\documentclass[11pt,a4paper]{article}
\usepackage{amssymb}
\usepackage{amsmath}
\usepackage{xcolor}
\usepackage{amsfonts}
\usepackage[margin=1in,footskip=0.25in]{geometry}

\newcommand{\be}{\begin{equation}}\newcommand{\ee}{\end{equation}}
\newcommand{\bea}{\begin{eqnarray}}\newcommand{\eea}{\end{eqnarray}}
\newcommand{\brr}{\begin{array}}\newcommand{\err}{\end{array}}
\newcommand{\bit}{\begin{itemize}}\newcommand{\eit}{\end{itemize}}
\newcommand{\ben}{\begin{enumerate}}\newcommand{\een}{\end{enumerate}}

\newcommand{\ba}{\begin{array}}
\newcommand{\ea}{\end{array}}

\def\lf{\left}

\def\ri{\right}
\def\al{\alpha}

\def\1{{_{1}}}\def\2{{_{2}}}

\def\noHe0{:\;\!\!\;\!\!:H_e(0):\;\!\!\;\!\!:}
\def\noHm0{:\;\!\!\;\!\!:H_\mu(0):\;\!\!\;\!\!:}

\def\lf{\left}

\def\ri{\right}

\def\al{\alpha}

\def\1{{_{1}}}\def\2{{_{2}}}

\begin{document}

\title{\textbf{Testing gravity with neutrinos: from classical to quantum regime}\\[5mm]
\small{\textbf{Essay written for the Gravity Research Foundation 2020 Awards for Essays on Gravitation.}}}

\author{Giuseppe Gaetano Luciano\thanks{gluciano@sa.infn.it}\\ Universit\`a degli Studi di Salerno \& INFN, Sezione di Napoli
\and Luciano Petruzziello\thanks{lupetruzziello@unisa.it, corresponding author}\\ Universit\`a degli Studi di Salerno \& INFN, Sezione di Napoli}

\vspace{4mm}

\date{\today}
\def\be{\begin{equation}}
\def\ee{\end{equation}}
\def\al{\alpha}
\def\bea{\begin{eqnarray}}
\def\eea{\end{eqnarray}}
\maketitle
\begin{abstract}
In this manuscript, we survey the main 
characteristics that provide neutrinos with the 
capability of being the perfect 
candidate to test gravity. 
A number of potentially resourceful scenarios 
is analyzed, with particular emphasis on how the
versatility of neutrinos
lends itself to understand the 
multifaceted nature of the gravitational interaction, 
both at classical and quantum scales.
As a common thread running through the 
two different regimes, 
we consider the
fundamental principles underpinning General Relativity
and its possible quantum extensions.
Finally, we discuss some open problems and
future perspectives. 
\end{abstract}

\vskip -1.0 truecm

\section*{Introduction}
Neutrinos are the most elusive elementary
particles in the Standard Model. 
Due to their extremely small mass and
zero electric charge, they are capable of passing
through ordinary matter with minimal interaction, 
representing a unique probe for investigating
physics at length scales ranging
from nuclei, to molecules and
galaxies.
Besides, the challenging search for direct evidences of the 
Cosmic Neutrino Background (CNB) may 
provide us with fundamental knowledge 
on the earliest stages of Universe's existence. 
Due to these peculiar features, 
neutrinos can thus be regarded as 
unparalleled information messengers
in many branches of physics. Among these, 
gravity theories and the related host of 
unsolved problems certainly 
represent one of the most demanding fields of research.

\medskip
The aim of this work is twofold:
\begin{itemize}
\item[(i)] on the one hand, neutrino physics is used as a test bench
for predictions of General Relativity (GR) and its cornerstones
at the classical and semi-classical level;

\item[(ii)] on the other hand, we discuss
how the above framework may potentially
unravel the unsettled riddles arising in the regime where 
General Relativity and 
Quantum Mechanics should coexist.  
\end{itemize}

\section*{Neutrino physics and classical gravity}
In the extended Standard Model, it is well-known that
neutrinos weakly interact in flavor states $|\nu_\alpha\rangle$
that are superpositions of mass states $|\nu_k\rangle$
according to\footnote{We shall work within a simplified two-flavor scenario
and in the approximation of relativistic neutrinos.}~\cite{pontecorvo}
\be\label{flavor}
|\nu_\alpha\rangle=\sum_{k=1,2}U_{\alpha k}(\theta)|\nu_k\rangle\,, \qquad \alpha=e,\mu\,,
\ee 
where $U_{\alpha k}$ is the
generic element of Pontecorvo matrix.

Mass states propagate freely. In Minkowski
spacetime, their evolution from a point $A(t_A,\vec{x}_A)$ to $B(t_B,\vec{x}_B)$ is governed by the phase factor $\varphi_k=E_k(t_B-t_A)-\vec{p}_k\cdot(\vec{x}_B-\vec{x}_A)$, where $E_{k}=\sqrt{m_k^2+|\vec{p}_k|^2}$ and $\vec{p}_k$
denote the energy and three-momentum of the $k^{th}$ state of mass $m_k$, respectively.
Accordingly, 
the phase shift $\varphi_0$ acquired by the mass eigenstates 
during the propagation leads to a non-vanishing flavor transition probability 
\be
\mathcal{P}_{\alpha\to\beta}=\sin^2(2\theta)\sin^2\lf(\frac{\varphi_0}{2}\ri),
\ee 
where
\be\label{phase}
\varphi_0\simeq\frac{\Delta m^2}{2E_\ell}L_p\,, \qquad \Delta m^2=m_2^2-m_1^2\,,
\ee
with $E_\ell$ being the (common) local energy of neutrinos 
and $L_p$ the proper distance they travel.

Inspired by the detection of a Newtonian gravitational phase in a neutron-based interferometry experiment~\cite{cow}, Stodolski~\cite{stodolski} first investigated GR effects on the wave functions of particles propagating in curved background. 
Should the analysis of flavor oscillations 
be performed for neutrinos
in the gravitational field of 
a source mass $M$, GR would then predict
for the phase shift~\cite{oscillation}
\be\label{sum}
\varphi=\varphi_0+\varphi_{GR}(M)\,.
\ee
In the simplest case of Schwarzschild
spacetime and in the weak-field limit, one has
\be\label{grphase}
\varphi_{GR}(M)\simeq\frac{\Delta m^2 L_p}{2E_\ell}\lf[\frac{GM}{r_B}-\frac{GM}{L_p}\ln\lf(\frac{r_B}{r_A}\ri)\ri].
\ee
Although very small in size, the correction~\eqref{sum}
may be in principle attainable in neutrino interferometry experiments~\cite{giunti}, 
thus allowing for a direct test of GR predictions
via neutrino oscillations in regimes of weak gravity. 
{However, describing the
background gravitational field by simply using the 
spherically-symmetric Schwarzschild solution is not satisfactory in most
situations. Astrophysical sources are expected to be rotating
as well as endowed with shape deformations leading to effects
which, in general, cannot be neglected. As a matter of fact,   
in order to render the above picture as realistic as possible, one should 
perform the analysis of neutrino oscillations 
in Kerr spacetime. A first step along this direction
has been taken in Ref.~\cite{konno}, where 
gravity corrections to the phase shift have been 
computed in the slowly rotating, weak-field approximation and
for the case of ultra-relativistic, spin-$1/2$ particles 
described by left-handed Weyl spinors. 
Further clues may come from the generalization 
of the above formalism to more exotic (quasi-spherical) geometries, 
where the assumption of a Kerr-like metric 
may lead to erroneous conclusions about 
the actual astrophysical processes that take place.
For instance, neutrino oscillations in the field
of rotating deformed neutron stars, white dwarfs and supermassive
stars can be reasonably described by the Hartle-Thorne~\cite{HT}
or Zipoy-Voorhees metrics~\cite{Zipoy}.}

The aforementioned analysis refers to 
vacuum flavor transitions and holds
in the weak-field approximation. With
proper refinements which embed 
matter enhancing effects (MSW effect)
and higher-order gravity corrections, 
it can be safely extended to a
variety of astrophysical environments. 
For instance, it is a well-established fact
that neutrinos play a crucial r\^ole 
in stellar collapses and formation of black holes 
and neutrons stars. Specifically, the theoretical models 
describing these phenomena~\cite{newyork} 
were proved to be accurate on the basis 
of the first-ever neutrino burst detection 
coming from the supernova SN1987A~\cite{supernova}. 
{In this regard, it must be pointed out that, according to Ref.~\cite{wolf},
the matter effects attributable to the high density of the collapsing stellar cores
were believed to inhibit neutrino oscillations. Under a similar
circumstance, the inevitable conclusion would be a permanent trapping of
neutrinos due to the absence of flavor transition, which instead could permit 
a leakage from the surrounding astrophysical environment. However, 
the matter-induced suppression of oscillations 
was subsequently reconsidered 
in Ref.~\cite{ahluw2}, showing that the relevance
of neutrino oscillations in regimes of strong-gravity
and, in particular, in supernova
explosions, could strongly depend on 
the distribution of space regions where matter effects are factually prominent.
Remarkable results supporting the key r\^ole
of neutrinos in driving the collapse and explosion of massive
stars have been recently summarized in Ref.~\cite{Mirizzi} (and therein), 
where it has been argued that, due to their weakly interacting nature,
these particles represent the only direct probe of the dynamics and thermodynamics at the center of a supernova. In particular, 
hydrodynamical simulations with
most sophisticated neutrino transport have proved 
to be necessary to calculate detailed signal properties,
which are required for the analysis of neutrino oscillations and 
neutrino-induced nucleosynthesis in supernovae, 
and for the potential detection of the Diffuse Supernova Neutrino Background (DSNB)  and of neutrinos from a future Galactic supernova.}

Furthermore, 
(heavy sterile) neutrinos are regarded 
as candidates for Dark Matter~\cite{dm} and 
as probes to reveal the mysterious nature of dark energy~\cite{Dvali}, which
potentially offers precious 
hints towards the resolution of GR
main puzzles.
{Among these, one of the most intriguing issues is represented by the
intrinsic importance of the torsion $T$, which always equals zero in the context of GR.
In spite of this, a number of works has been developed with the 
assumption $T\neq0$, which gives rise to the so-called ``torsion gravity'' models~\cite{tg}.
The interest in such models can be ascribed to their capability
of avoiding singularities, both at the quantum~\cite{pop1} and the cosmological~\cite{pop2} level.
Even in this framework, the 
study of neutrino oscillations may 
have non-trivial implications. As shown in Refs.~\cite{neutorsion}, 
the shape of the flavor transition probability can  
provide clues on the existence of a non-vanishing torsion, thus allowing
to discriminate between the standard GR scenario and torsion gravity. 
In a two-flavor configuration, these effects (which are of the order of Planck scale) become manifest only when the superposed mass eigenstates
have opposite spin, otherwise no discrepancy with standard GR results
arises at all.} 

On the other hand,  
neutrino physics can provide valuable 
pieces of information about 
the principles underlying 
GR and other gravitational models.
Indeed, from the first 
available data on astrophysical neutrinos,
such particles have been constantly associated 
to the violation of the weak equivalence principle~\cite{wep1,wep2}. 
Additionally, from a more theoretical perspective, 
a similar 
scenario is encountered within the framework of 
exotic geometries and extended 
models of gravity~\cite{examples}.
All of these evidences lead to the awareness 
that the equivalence principle should be somehow modified when 
passing from classical to quantum regimes,  
as preliminarily pointed out in Ref.~\cite{brukner}. 

\section*{Neutrino physics: from semiclassical to quantum gravity}
Along with the equivalence principle, 
general covariance represents another fundamental pillar 
of GR. Contrary to the former, however, 
such a principle still
underpins most of the attempts of extending GR  
made so far. For instance, the generalization 
of Quantum Field Theory 
(QFT) to curved background is by construction generally covariant. 
Although this model only provides
a semiclassical description of gravitational interaction, 
a plethora of its predictions 
are subjects of active investigation. 
Among these, 
the Hawking-Unruh radiation is certainly 
the most eloquent footprint of a possible 
non-classical nature of gravity.
In this regard, many studies
predict that neutrinos
expelled during black hole evaporation may non-trivially 
affect the emitted power and the lifetime of 
the source~\cite{bam}, with phenomenological  
consequences which may be relevant for 
ruling out primordial black holes as Dark Matter candidates~\cite{pbh}. 
Furthermore, processes involving the production and/or 
absorption of neutrinos can be used as a theoretical tool for testing the 
existence of the Unruh effect as a consequence of the general covariance of QFT~\cite{matsas,Ahluwalia:2015kxa,Blas,Cozzella:2018qew}, as well as deviations of the Hawking-Unruh 
spectrum from a purely thermal behavior~\cite{Non-therm}.

\smallskip
In connection with the issue of 
fundamental principles, 
let us observe that 
another stimulating 
link between the ``neutrino'' and ``gravity'' worlds is
provided by string theory's prediction of 
the existence of a minimum length at Planck scale $\lambda_P\simeq10^{-35}\,$m, 
in compliance with the possible emergence 
of a discrete structure of spacetime. 
Implications of this requirement 
are extremely non-trivial, as they would affect most of
the basic principles of modern physics, 
such as (local) Lorentz invariance and  
Heisenberg Uncertainty Principle (HUP). 
In all the cases, 
theoretical and experimental investigations  
involving neutrinos 
may shed light on such peculiar features. 
Specifically, detailed 
studies on Lorentz violation in neutrino oscillations 
have been proposed in Ref.~\cite{kost}. On the other hand, 
signatures of Planck-scale corrections
to Pontecorvo oscillation formula
have been addressed in Ref.~\cite{gup} using a
generalized commutator (GUP) of the form
\begin{equation}
[\hat{x},\hat{p}]\,=\,i\left(1+f(|p|^2)\right),
\end{equation}
where $p$ is the characteristic momentum of the physical
system and $f(|p|^2)\rightarrow0$ at energies far from Planck scale, 
so as to recover the standard quantum mechanical
framework. {In this perspective, a suggestive
prediction has been conjectured in Ref.~\cite{Ahluwalia:2000iw} on the
basis of a GUP-modified de Broglie formula describing the 
wave-particle duality in the Planck regime, that is
\be
\label{GUPlam}
\lambda_{dB}\sim\frac{1}{p}\,\,\,\,\overset{{GUP}}{\longrightarrow}\,\,\,\, \lambda\sim\frac{\lambda_P}{\tan^{-1}(\lambda_P/\lambda_{dB})}\left\{
                \begin{array}{ll}
                  \rightarrow \lambda_{dB}\,\,\,\,\, \mathrm{for\,\, low-energy}\\
                  \rightarrow \lambda_P\,\,\,\,\hspace{2.2mm} \mathrm{at\,\, Planck\,\,scale}
                \end{array}
              \right..
\ee 
In fact, by attributing
the origin of low-energy neutrino oscillations to the
different de Broglie oscillation lengths associated with each mass
eigenstate, it has been argued that the phenomenon of flavor
changing may be \emph{freezed} at Planck scale, owing to the saturation of Eq.~\eqref{GUPlam} for all mass eigenstates. 
Nevertheless, due to the number of still open theoretical questions and the
lack of experimental guidance at Planck energy, a definitive conclusion about
the actual occurrence of the freezing of oscillations in all neutrino frameworks (equal energy, equal velocity or wave packet approaches) has not yet been reached.}

\smallskip
Beyond theoretical conjectures, 
a more phenomenological investigation of
non-standard features of gravity
is related to the challenging detection of the
Cosmic Neutrino Background~\cite{z, tritium}, 
whose existence is 
supported only by
strong indirect evidences to date~\cite{data}.
Since relic neutrinos
decoupled from matter few seconds 
after the Big Bang, it is possible to extract 
a great amount of data on the primordial features of the Universe 
from them. In that stage, quantum and gravitational effects are 
expected to be comparably important, 
thus promoting such particles as unique
witnesses of ``exotic'' gravity regimes
that can no longer be reproduced in laboratory.

\smallskip
Finally, even though 
difficulties in testing the quantum
nature of gravity seemed to 
relegate models such as QFT on curved background, string theory and loop quantum gravity   
to merely speculative formalisms 
until a few years ago, a promising way out
has been recently offered by 
a series of experiments aiming  at
characterizing gravity as a \emph{quantum coherent mediator}.
The idea (which traces back to Feynman) 
is to consider
two test masses prepared so as to 
exclude all types of perturbations from the environment
and among each other, 
except for the mutual gravitational interaction. 
Then, if at a certain time 
a non-vanishing entanglement is measured
between them,
the only reason for this
would be the exchange of a graviton, which 
would certify a sort of gravity quantumness~\cite{qtest}. 
In principle, a similar reasoning 
can be carried out also for superpositions of mass states~\cite{nugrav} and, 
thus, for neutrinos. The advantage
of using these particles is that 
they are only affected by the weak interaction 
and gravity, which significantly simplifies the
realization of the experimental setup.

\section*{Future perspectives}
In this work we have sorted through
some of the scenarios where
neutrinos act as a probe for testing
gravity, both at classical and quantum scales.
Even though most of them
are genuine smoking guns since long time, 
several others are being proposed only
in recent years.
Let us mention some of the most promising ones:

\begin{itemize}
\item[(i)] gravitational waves (GW)
have been shown to non-trivially 
affect neutrino spin and flavor oscillations.
In particular, the case of neutrino interaction
with stochastic GWs emitted by coalescing supermassive black holes
has been discussed~\cite{dvornikov}.
The question thus arises
as to whether this mechanism can be 
exploited to gain information
about physics of the GW emitting source
through the detection of 
neutrinos undergoing oscillations in such a gravitational background;

\item[(ii)] in the last Section, we
have discussed implications
of some proposed theories of quantum gravity
for neutrino oscillations within 
the GUP framework. Note that, although 
the induced corrections are strongly suppressed
by Planck energy, 
they may be experimentally detectable
for ultra-high-energy cosmogenic neutrinos~\cite{wk}. 
Therefore, we expect that 
the next-generation neutrino detectors
may provide significant contributions
in this direction.
\end{itemize}
Clearly, finding definite solutions to the
above problems  
and framing them within a unified 
picture is a demanding, but
at the same time intriguing, task. 
More investigation is inevitably
required along this line.

\end{document}